# Statistical Distance-Guided Unsupervised Domain Adaptation for Automated Multi-Class Cardiovascular Magnetic Resonance Image Quality Assessment

Shahabedin Nabavi[1], Kian Anvari Hamedani[1], Mohsen Ebrahimi Moghaddam[1], Ahmad Ali Abin[1], Alejandro F. Frangi [2,3,4]

1- Faculty of Computer Science and Engineering, Shahid Beheshti University, Tehran, Iran.
2- Division of Informatics, Imaging and Data Sciences, Schools of Computer Science and Health Sciences, The University of Manchester, Manchester, UK.
3- Medical Imaging Research Center (MIRC), Electrical Engineering and Cardiovascular Sciences Departments, KU Leuven, Leuven, Belgium.
4- Alan Turing Institute, London, UK.

**Corresponding Author:** Mohsen Ebrahimi Moghaddam

**Address:** Faculty of Computer Science and Engineering, Shahid Beheshti University, Tehran, Iran.

**Email:** m_moghadam@sbu.ac.ir

**Phone:** +98 912 140 5308

**Statements and Declarations**

There are no conflicts of interest to declare. No funding was received for conducting this study.

**Data Availability Statement:** The datasets are available at

- CMRxRecon2023: https://cmrxrecon.github.io/
- Automated Cardiac Diagnosis Challenge (ACDC): https://www.creatis.insa-lyon.fr/Challenge/acdc/databases.html
- CMRxMotion: http://cmr.miccai.cloud/
- UK Biobank is available by completing a formal access application: https://www.ukbiobank.ac.uk/enable-your-research/apply-for-access

**Code Availability Statement:** The codes are available at

https://github.com/kiananvari/Statistical-Distance-Guided-Unsupervised-Domain-Adaptation-for-Automated-Multi-Class-CMR-IQA

**Abstract**

This study proposes an attention-based statistical distance-guided unsupervised domain adaptation model for multi-class cardiovascular magnetic resonance (CMR) image quality assessment. The proposed model consists of a feature extractor, a label predictor and a statistical distance estimator. An annotated dataset as the source set and an unlabeled dataset as the target set with different statistical distributions are considered inputs. The statistical distance estimator approximates the Wasserstein distance between the extracted feature vectors from the source and target data in a mini-batch. The label predictor predicts data labels of source data and uses a combinational loss function for training, which includes cross entropy and centre loss functions plus the estimated value of the distance estimator. Four datasets, including imaging and k-space data, were used to evaluate the proposed model in identifying four common CMR imaging artefacts: respiratory and cardiac motions, Gibbs ringing and Aliasing. The results of the extensive experiments showed that the proposed model, both in image and k-space analysis, has an acceptable performance in covering the domain shift between the source and target sets. The model explainability evaluations and the ablation studies confirmed the proper functioning and effectiveness of all the model's modules. The proposed model outperformed the previous studies regarding performance and the number of examined artefacts. The proposed model can be used for CMR post-imaging quality control or in large-scale cohort studies for image and k-space quality assessment due to the appropriate performance in domain shift coverage without a tedious data-labelling process.

## 1- Introduction

Cardiovascular magnetic resonance (CMR) imaging has become a gold standard in diagnosing several cardiovascular diseases due to its non-invasive nature and anatomical/functional diagnostic abilities [1, 2]. Imaging artefacts have a side effect on the diagnostic and quantitative analysis, and can derail the treatment process. Although novel MR protocols and cutting-edge equipment have reduced the incidence of artefacts, the occurrence of those is still inevitable. This issue emphasises the necessity of developing automatic methods in artefact detection to reduce costs and improve services to patients in busy imaging centres. Providing automatic methods based on deep learning is associated with challenges such as the availability of bulk annotated datasets, developing a no-reference approach, and the domain shift problem. Deep neural networks have a data-hungry nature in such a way that there is a need for a large annotated dataset to train them [3], which complicates the task due to the costly and laborious process of data labelling in medical applications [4]. Due to the lack of access to reference images for medical image quality assessment, the presented approach should be a no-reference method [5]. Besides, a domain shift between training data and data used in the practical application of deep learning approaches drops the performance of these methods [6]. The domain shift between different datasets is one of the leading challenges of using deep learning models for clinical applications.

Several studies have developed automated methods to detect CMR image artefacts [7-12] and control the complete coverage of the left ventricle [13-15]. Few methods have been previously proposed for quality control of these images considering approaches such as domain adaptation [16], domain adaptive knowledge distillation [17], and meta-learning [18].

This study proposes an attention-based statistical distance-guided unsupervised domain adaptation model for multi-class CMR image quality assessment. This model can identify four common CMR artefacts, including respiratory motion, cardiac motion, Gibbs ringing and aliasing, by receiving the image or k-space from an annotated source dataset and an unlabeled target dataset with different statistical distributions. The model can control the quality of input images/k-spaces from 2, 3 or 4-chamber long-axis or short-axis views, so the model has more generality from this point of view. Besides, the proposed model can jointly cover the domain shift between source and target sets and extract discriminative features without requiring data labelling for the target set. Thus, it can improve the performance on the target set through the knowledge of the annotated source set.

## 2- Materials and Methods

In this section, the datasets used in the study are first described. Then, the proposed model for identifying CMR image artefacts is presented. Finally, the training details and experimental setups are mentioned to explain the training process, the defined experiments and evaluation metrics.

### 2-1- Dataset description and Preprocessing

Four datasets, including UK Biobank (UKBB) [19], the automated cardiac diagnosis challenge (ACDC) [20], the cardiac MRI reconstruction challenge (CMRxRecon) in 2023 [21], and the extreme cardiac MRI analysis challenge under respiratory motion (CMRxMotion), were used to prepare training and testing data for evaluating the proposed method. These datasets have a different domain shift or statistical distribution compared to each other (See Figure 1). Based on subjective evaluations, several patients' artefact-free CMR imaging data were selected from the UKBB, ACDC, and CMRxRecon datasets. The patient-out method was leveraged to prepare training and test sets to ensure the correctness of the results. The information about the training and test sets is given in Table 1.

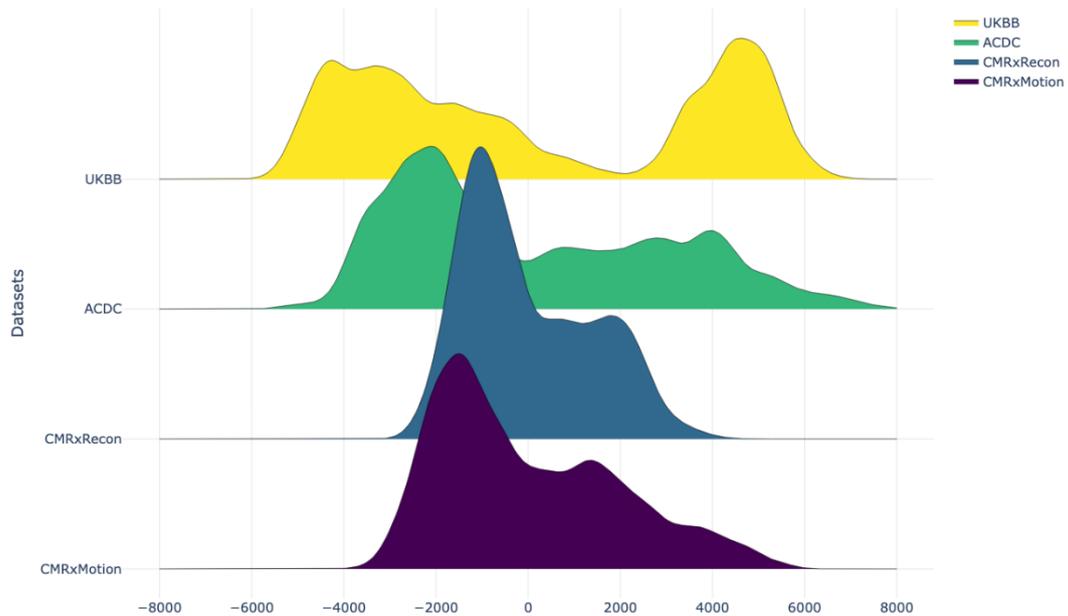

**Figure 1. Distribution visualisation of samples regarding the UKBB, ACDC, CMRxRecon, and CMRxMotion datasets.**

**Table 1. Information about the size of the training and testing sets used in the current study.**

|  | # of Patients | | # of 2D slices | |
|---|---|---|---|---|
|  | **Training set** | **Testing set** | **Training set** | **Testing set** |
| UKBB | 120 | 20 | 30,000 (6,000 per class) | 5,000 (1,000 per class) |
| ACDC | 25 | 6 | 30,000 (6,000 per class) | 5,000 (1,000 per class) |
| CMRxRecon | 25 | 5 | 30,000 (6,000 per class) | 5,000 (1,000 per class) |
| CMRxMotion | 16 | 4 | 1,210 (605 per class) | 300 (150 per class) |

CMR images may be associated with artefacts due to hardware and software restrictions. These artefacts can seriously affect image quality and, thus, the ability to make a proper diagnosis. According to the published statistics, motion and Gibbs artefacts are the most common in magnetic resonance imaging [22]. Specifically, respiratory and cardiac motion, Gibbs ringing, and aliasing artefacts are the most common in CMR imaging [23]. For this reason, we investigated these types of artefacts in this study.

Subjectively determining the artefact in the CMR images is a time-consuming and laborious task that requires several human observers. Since this study needs corrupted images, we used the k-space degradation methods to generate synthetic but realistic corrupted imaging data. Long- and short-axis cine CMR images acquired using Cartesian sampling were subjected to add respiratory and cardiac motion, aliasing, and Gibbs ringing artefacts. For generating corrupted CMR images based on the k-space manipulation methods, we used the model proposed in [8, 24] for respiratory motions, [8] for cardiac motions, the signal truncation method based on ideal low-pass filtering [25, 26] for Gibbs ringing, and the k-space subsampling presented by [27] for aliasing artefact. The intensity of the added artefacts was considered randomly to increase the proposed method's generality. An example of degraded images with the mentioned methods is shown in Figure 2.

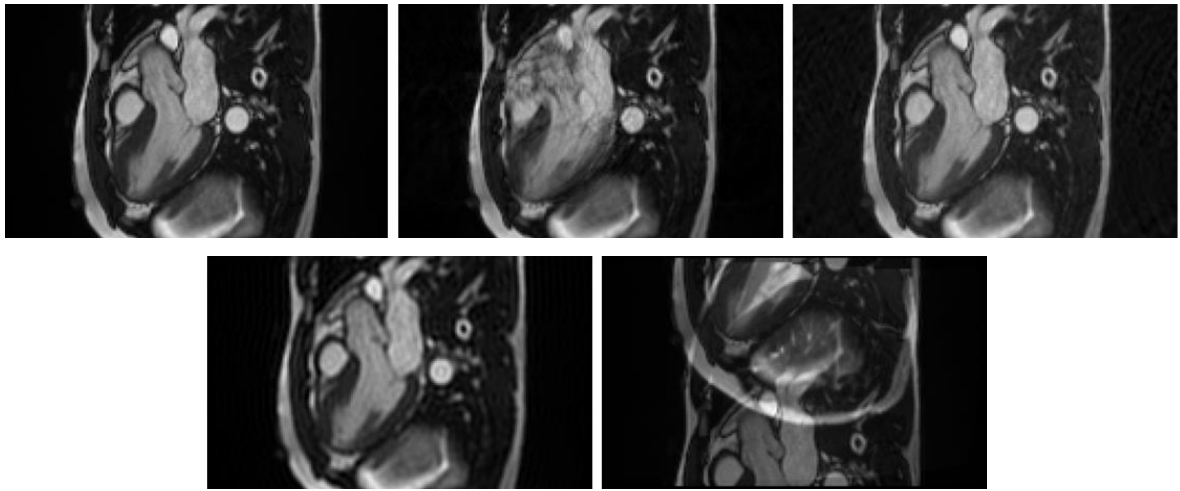

**Figure 2. An example of corrupted images with k-space manipulation methods. The first row (from left to right) is the artefact-free reference image, the degraded image with the cardiac and respiratory motion artefacts. The second row (from left to right) is the degraded image with Gibbs ringing and aliasing artefacts.**

**2-2- The Proposed Method**

An unsupervised adversarial domain adaptation method is proposed for CMR image quality assessment in the presence of domain shift between datasets. Relying on the customisation of loss functions, this method simultaneously seeks to achieve two goals, including the distribution adaptation of the source and target datasets and increasing the discriminability of samples from different classes. The overview of the proposed method is shown in Figure 3.

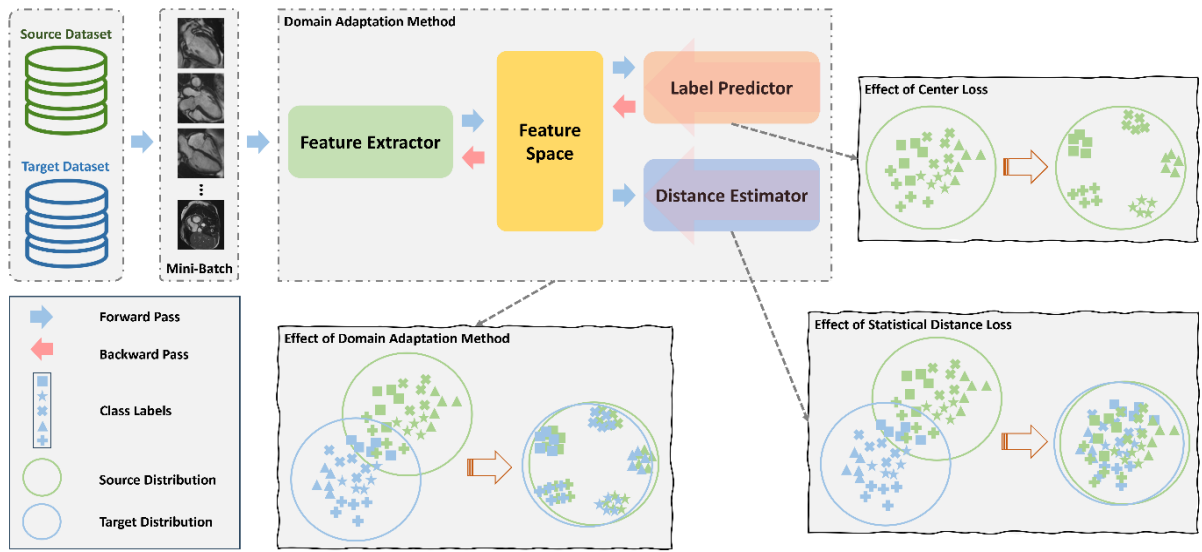

Figure 3. The overview of the proposed domain adaptation method.

The proposed method considers two datasets with domain shift as input data. The source dataset is annotated, while the target dataset has no class label for samples. The source and target data are shuffled to create mini-batches necessary for training the model. The mini-batch's data is fed to a convolutional neural network for feature extraction. The extracted feature vectors for the source data are given to both label predictor and distance estimator networks, while the target data is only given to the distance estimator. In the proposed method, the label predictor should be able to reduce the intra-class distance and increase the inter-class distance to improve the separability of the different classes. The distance estimator tries to approximate the statistical distance between the source and target data distributions in each mini-batch. The estimated distance is used in the combinational loss function to cover the domain shift. These two networks jointly result in extracting domain-independent features with high separability for each class. As a result, the model's performance is improved when faced with unlabeled target data. Algorithm 1 describes the proposed method.

**Algorithm 1** Statistical Distance Guided Unsupervised Domain Adaptation

**Require:**
    ▷ $\mathcal{D}_S = \{(x_i, y_i)\}_{i=1}^N$ is an annotated source dataset
    ▷ $\mathcal{D}_T = \{x'_j\}_{j=1}^N$ is an unlabelled target dataset
    ▷ $\theta$ : Initial parameters of the feature extractor
    ▷ $p$ : Initial parameters of the label predictor
    ▷ $e$ : Initial parameters of the statistical distance estimator
    ▷ $\lambda, \gamma$ : Adjustment coefficients

**Ensure:**
    ▷ $\theta'$ : Final parameters of the feature extractor
    ▷ $p'$ : Final parameters of the label predictor

1: **for** $iteration = 1, 2, ...$ **do**
2:     Select a mini-batch from $D_S$ and $D_T$
3:     **for** All $(x_i, y_i)$ and $x'_j \in$ mini-batch **do**
4:         $z_i^S \leftarrow f_\theta(x_i)$
5:         $z_j^T \leftarrow f_\theta(x'_j)$
6:         **for** $step = 1, 2, ...$ **do**
7:             $z_k^* \leftarrow$ Sampling along the straight line between pairs of $z_i^S$ and $z_j^T$
8:             $z \leftarrow concat(z_i^S, z_j^T, z_k^*)$
9:             $\mathcal{L}_{grad\_penalty}(z) = (\|\nabla_z f_e(z)\|_2 - 1)^2$
10:            $\mathcal{L}_{dist\_estimator}(z_i^S, z_j^T) = \frac{1}{N_S}\sum_i z_i^S - \frac{1}{N_T}\sum_j z_j^T$
11:            $e \leftarrow$ Adam_Optimizer $([\mathcal{L}_{dist\_estimator}(z_i^S, z_j^T) - \lambda \mathcal{L}_{grad\_penalty}(z)], e)$
12:         **end for**
13:         $pred \leftarrow f_p(z_i^S)$
14:         $\mathcal{L}_{CE}(pred, y_i) = -\frac{1}{N_S}\sum_{i=1}^{N_S}(y_i.\log(pred) + (1 - y_i).\log(1 - pred))$
15:         $\mathcal{V}_i^S \leftarrow$ The penultimate layer output of the label predictor for $z_i^S$
16:         $c_l \leftarrow$ Averaging the feature vectors $\mathcal{V}_i^S$ for samples of class label $l$
17:         $\mathcal{L}_{center}(z_i^S, c_l) = \frac{1}{2}\sum_{i,l} \|z_i^S - c_l\|_2^2$
18:         $\gamma = \frac{iteration}{number\ of\ iterations}$
19:         $\mathcal{L}_{stat\_dist}(z_i^S, z_j^T) = \frac{1}{N_S}\sum_i f_e(z_i^S) - \frac{1}{N_T}\sum_j f_e(z_j^T)$
20:         $\theta', p' \leftarrow$ Adam_Optimizer $([\mathcal{L}_{CE} + \mathcal{L}_{center} + \gamma \mathcal{L}_{stat\_dist}], \theta, p)$
21:     **end for**
22: **end for**
23: **return** $\theta', p'$     ▷ Final Parameters of the Domain Adapted Network

As it appears from the algorithm, the distance estimator network uses the Wasserstein metric [28] to estimate the distance between the representation distribution of the source and target sets. For this purpose, this network receives the representations of the source and target data in each mini-batch and it calculates the Wasserstein distance based on equation 1 to update the distance estimator network:

$$\mathcal{L}_{dist\_estimator}(z_i^S, z_j^T) = \frac{1}{N_S}\sum_i z_i^S - \frac{1}{N_T}\sum_j z_j^T \tag{1}$$

where $z_i^S$ and $z_j^T$ are respectively the representations obtained from the feature extractor for source and target samples of the respective mini-batch. $N_S$ and $N_T$ are the numbers of samples belonging to the source and target in the mini-batch, respectively.

It is necessary to apply the Lipschitz constraint on the distance estimator network to obtain the empirical Wasserstein metric. Gulrajani et al. [29] have suggested that instead of using weight clipping, which results in gradient vanishing or exploding problems, a gradient penalty should be used to update the parameters of the distance estimator network to apply the constraint. This gradient penalty is calculated by equation 2.

$$\mathcal{L}_{grad\_penalty}(z) = (\|\nabla_z f_e(z)\|_2 - 1)^2 \quad (2)$$

where $z$ contains pairs of representations of source and target samples as well as random points along the straight line between pairs of representations.

Thus, the combinational loss of equation 3 is used to update the parameters of the distance estimator network. To obtain more optimization, the training process of this network is done in several iterations to achieve more accurate distances.

$$\mathcal{L}_e = \mathcal{L}_{dist\_estimator}(z_i^S, z_j^T) - \lambda \mathcal{L}_{grad\_penalty}(z) \quad (3)$$

where $\lambda$ is an adjustment coefficient.

The training of the label predictor network is done using the combinational loss function of equation 4. In this equation, the cross entropy loss is utilised to estimate the classification error. Since we are not only looking for separable features and the discriminability of features is more important to achieve better separability in target samples, the centre loss [30] is also calculated as a part of the final loss function. It is also necessary to obtain the statistical distance for all mini-batch data by the distance estimator network and add it as the third term of the combinational loss function. Thus, the model can cover the domain shift between the source and target data distributions and open the way for better classification of the target data by obtaining discriminative features using the combinational loss function.

$$\mathcal{L}_p = \mathcal{L}_{CE} + \mathcal{L}_{center} + \gamma \mathcal{L}_{stat\_dist} \quad (4)$$

where $\gamma$ is an adjustment coefficient to control the effectiveness of the statistical distance term in the combinational loss function of equation 4.

### 2-3- Details of Model Training and Evaluation

The proposed model was trained and evaluated in the image and k spaces in extensive experiments. In all these experiments, the number of data in all classes is balanced to avoid bias. ResNet34 [31] was used as a feature extractor to train the proposed model. The Adam optimiser [32] was utilised in the training process, considering a learning rate of 0.01 with a step size of 20 and a decay of 0.5. The number of epochs was 50, and the batch size was 512, consisting of 256 samples from the source set and 256 from the target set. The 5-fold cross-validation in the patient-out manner was leveraged to evaluate the model. Accuracy, precision, recall, F1-score, and area under the receiver operating characteristic curve (AUC) [33] were considered to evaluate the proposed model. The adjustment coefficient $\lambda$ was set to 10 based on ablation studies. Besides, the structure of the label predictor and attention-based distance estimator networks is shown in Figure 4.

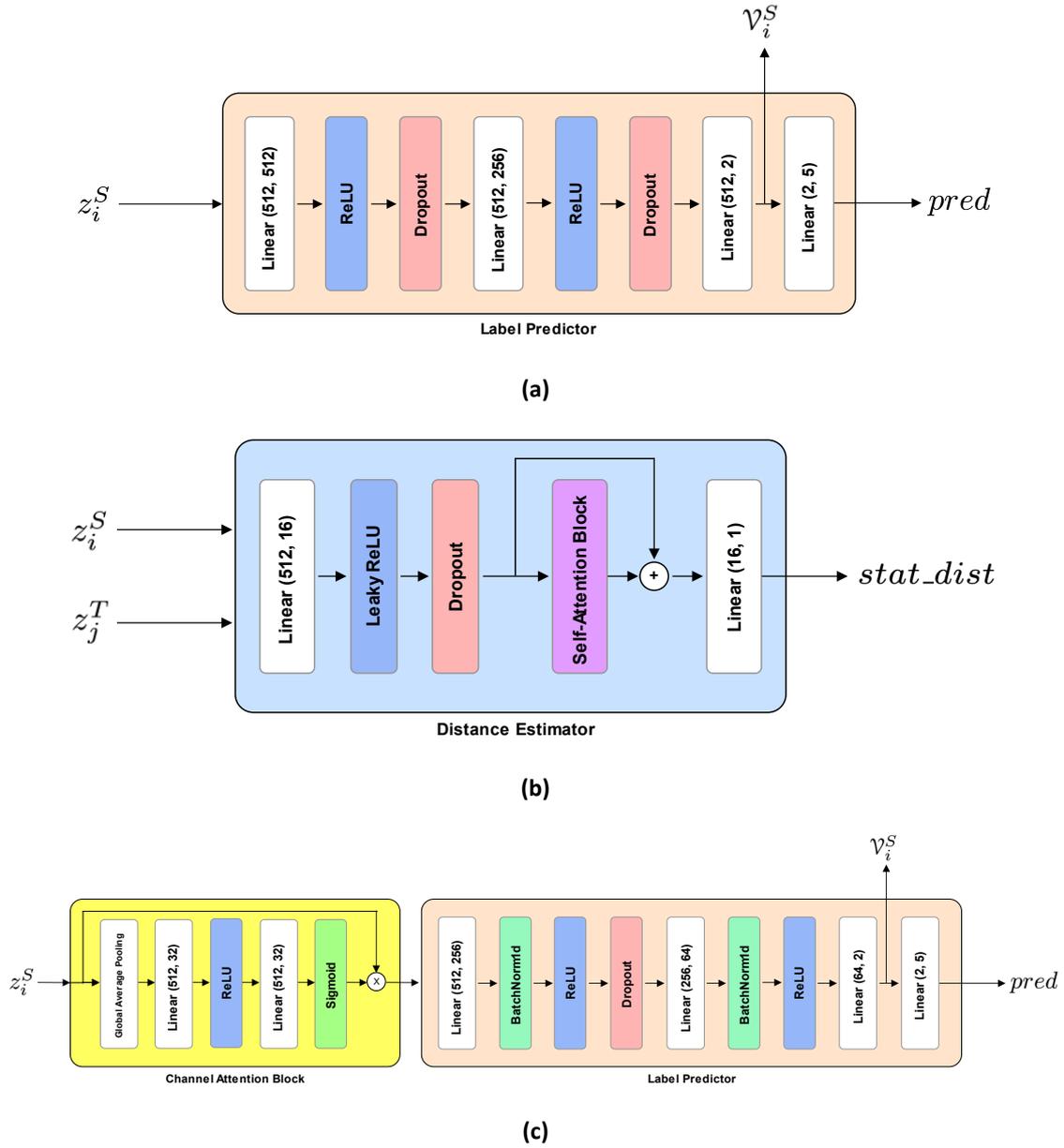

**Figure 4.** The network structures of (a) label predictor in the spatial domain, (b) distance estimator, and (c) label predictor for k-space analysis.

## 3- Results

To demonstrate the effectiveness of the proposed model, it is necessary to conduct experiments in the following three modes to reveal the coverage of the domain gap by the model.

- Training the model as source-only and testing it using target data (Train on source/Test on target). In this mode, because the target data for the model is unseen, we have the lowest values of the metrics.
- Training the model on a part of the target data and testing it on the rest of the target data (Train on target/Test on target). In this mode, due to the supervised training of the model on the target data, we have the maximum values of the metrics.

- Applying the proposed statistical distance-guided unsupervised domain adaptation model. In this mode, we expect the metrics' values to be approached to the maximum (mode 2) in an unsupervised manner, in other words, the model to be able to cover the domain gap.

## 3-1- Results of the Proposed Model in Spatial Domain

Considering the ACDC dataset as the source set and other datasets as the target set, the model evaluation results are presented in Table 2 in the three modes introduced. As can be seen from these results in the spatial domain, the proposed model has been able to cover the domain gap between different datasets with different distributions to a large extent. For a better understanding of the extent of domain gap coverage in the different experiments, the graph of Figure S1 is drawn based on the obtained accuracies (See supplementary materials). Besides, to evaluate the effectiveness of the proposed model in the five different classes studied, the confusion matrices before and after applying the proposed model are shown in Figure S2 (See supplementary materials).

Table 2. The results of the proposed method in the spatial domain. Results are based on patient-out 5-fold cross-validation ($Mean \pm STD$).

| Target | Metric | Accuracy | Precision | Recall | F1-Score | AUC |
|---|---|---|---|---|---|---|
| CMRxRecon | Train on source/Test on target | 43.52 ± 6.32 | 45.76 ± 14.21 | 43.52 ± 6.32 | 37.81 ± 6.54 | 74.00 ± 3.94 |
| | Using the proposed model | 84.92 ± 2.53 | 85.16 ± 2.31 | 84.92 ± 2.53 | 82.62 ± 2.38 | 94.88 ± 1.80 |
| | Train on target/Test on target | 90.49 ± 6.22 | 92.29 ± 5.32 | 90.49 ± 6.22 | 90.30. ± 6.36 | 98.69 ± 0.87 |
| UKBB | Train on source/Test on target | 46.96 ± 4.29 | 57.58 ± 9.51 | 46.96 ± 4.29 | 41.78 ± 4.15 | 76.78 ± 5.77 |
| | Using the proposed model | 78.33 ± 1.72 | 80.29 ± 2.76 | 78.33 ± 1.72 | 77.40 ± 4.56 | 87.58 ± 3.03 |
| | Train on target/Test on target | 80.29 ± 2.91 | 82.50 ± 3.65 | 80.29 ± 2.91 | 78.66 ± 3.66 | 95.97 ± 0.78 |
| CMRxMotion | Train on source/Test on target | 40.88 ± 6.52 | 35.68 ± 10.50 | 40.88 ± 6.52 | 30.66 ± 7.10 | 73.97 ± 5.18 |
| | Using the proposed model | 70.44 ± 2.91 | 77.41 ± 0.86 | 70.44 ± 2.91 | 68.17 ± 4.48 | 94.91 ± 1.34 |
| | Train on target/Test on target | 84.26 ± 3.99 | 85.64 ± 6.08 | 84.26 ± 3.99 | 82.93 ± 5.15 | 96.39 ± 1.91 |

The effectiveness of the proposed model based on visualization of the feature space during training iterations is shown in Figure 5 using the t-SNE [34] method. The first row of Figure 5 indicates that the proposed model has been able to adapt the data distribution of the source and target sets. Furthermore, in the second and third rows of this figure, the process of discriminating five different classes from each other can be seen on the source and target data, respectively.

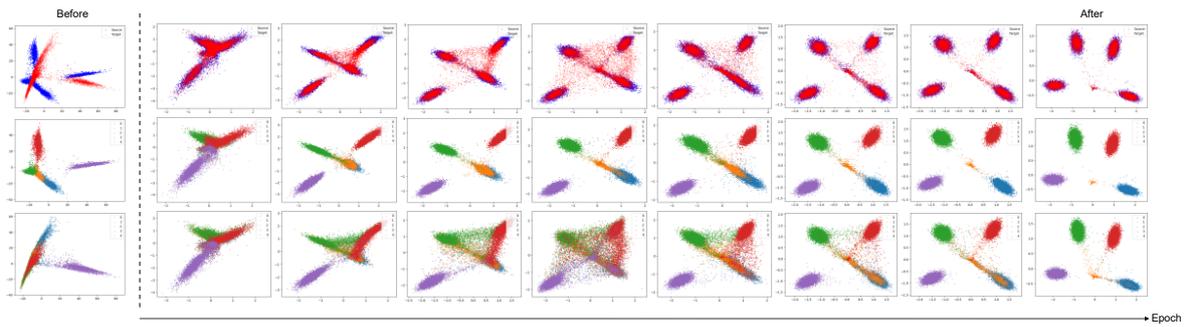

**Figure 5. The effect of the proposed model on the domain adaptability and the feature discriminability of different classes during the training process.**

In this study, the explainability of the proposed model was also examined. The Grad-CAM approach [35] has been leveraged to investigate the explainability of the proposed model. Figure 6 presents an example of the model's focus when identifying each of the five classes. These visual explanations indicate that the model is focused on the appropriate part of the image to determine the desired class.

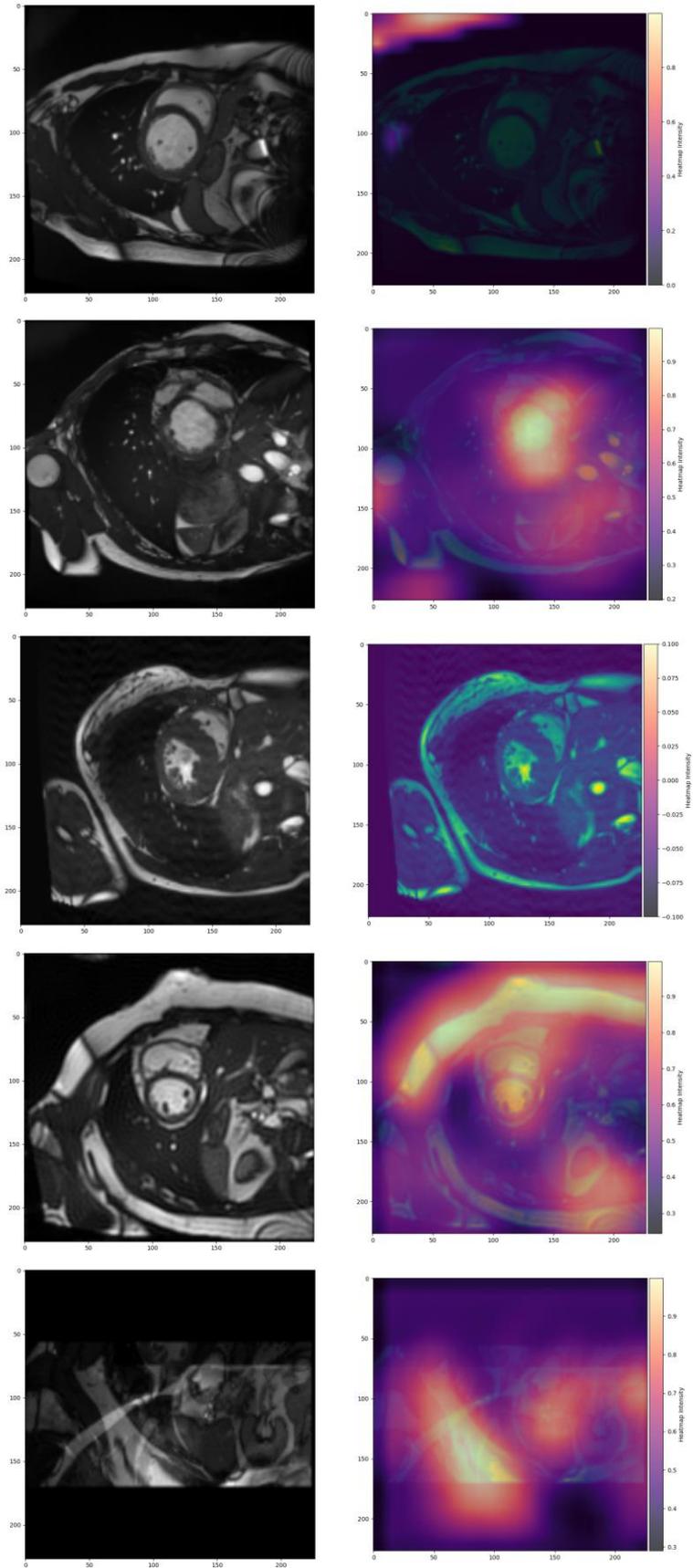

**Figure 6. The visual explanations regarding the decisions of the proposed model for (in order from the first row) artefact-free, cardiac motion, respiratory motion, Gibbs and Aliasing classes.**

## 3-2- Results of the Proposed Model in k-space

The proposed model for detecting artefacts based on receiving k-space instead of image was also evaluated. For this purpose, the model's input consisted of two stacked matrices containing real and imaginary parts of k-space. To achieve better results, the structure of the label predictor network was modified to evaluate the k-space data (Figure 4(c)). The results of using the proposed model in k-space, considering the CMRxRecon dataset as the source set and UKBB as the target set, are shown in Table 3.

**Table 3: The results of using the proposed model on the k-space data.**

| Dataset \ Metric | | Accuracy | Precision | Recall | F1-Score | AUC |
|---|---|---|---|---|---|---|
| **Source: CMRxRecon** **Target: UKBB** | Train on source/Test on target | 44.24 ± 5.73 | 37.93 ± 8.57 | 44.24 ± 5.73 | 36.38 ± 5.57 | 76.99 ± 4.61 |
| | Using the proposed model | 66.24 ± 5.09 | 74.24 ± 2.60 | 66.24 ± 5.09 | 61.58 ± 6.37 | 88.04 ± 3.35 |
| | Train on target/Test on target | 70.91 ± 6.41 | 74.38 ± 1.11 | 70.91 ± 6.41 | 68.25 ± 8.49 | 90.32 ± 1.62 |

## 3-3- Ablation Studies

A series of ablation studies were conducted to investigate the effectiveness of various components of the proposed model. These investigations were done in the spatial domain by considering the ACDC dataset as the source and CMRxRecon as the target set. The results of ablation studies are tabulated in Table 4.

**Table 4. The results of ablation studies. Results are based on patient-out 5-fold cross-validation ($Mean \pm STD$).**

| | Accuracy | Precision | Recall | F1-Score | AUC |
|---|---|---|---|---|---|
| **Train on source/Test on target** | 43.52 ± 6.32 | 45.76 ± 14.21 | 43.52 ± 6.32 | 37.81 ± 6.54 | 74.00 ± 3.94 |
| **Baseline Model + Center Loss** | 64.54 ± 2.27 | 74.67 ± 1.21 | 64.54 ± 2.27 | 60.78 ± 2.94 | 82.51 ± 1.98 |
| **Baseline Model + Wasserstein Distance Loss** | 71.50 ± 2.72 | 77.48 ± 0.77 | 71.50 ± 2.72 | 69.90 ± 4.38 | 91.38 ± 2.16 |
| **Baseline Model + Wasserstein Distance Loss (Attention-based Distance Estimator)** | 79.72 ± 0.68 | 81.9 ± 1.10 | 79.72 ± 0.68 | 78.95 ± 1.03 | 92.63 ± 1.65 |
| **Baseline Model + Center Loss + Wasserstein Distance Loss (Attention-based Distance Estimator) (Without Adjustment Coefficient $\gamma$)** | 82.04 ± 1.46 | 84.21 ± 2.39 | 82.04 ± 1.46 | 80.43 ± 1.43 | 93.94 ± 0.65 |
| **The Proposed Model** | **84.92 ± 2.53** | **86.16 ± 2.31** | **84.92 ± 2.53** | **82.62 ± 2.38** | **94.88 ± 1.80** |

### 3-4- The Proposed Model vs. Other Related Studies

In Table 5, the proposed model is compared with previous related studies. Studies have been compared from the perspective of distortions, domain adaptability, and learning type in this table. For fairness in these comparisons, the ACDC dataset was used as the source and CMRxRecon as the target set. Accuracy was not reported for studies that did not have domain adaptability, due to differences in the number of distortions examined.

Table 5. Comparison of the proposed model with related studies.

| Study | Modality | | Type of Distortion | | | | Domain Adaptability | Learning Type | Accuracy |
|---|---|---|---|---|---|---|---|---|---|
| | LAX | SAX | CM | RM | GB | AL | | | |
| Tarroni et al. [10] | - | ✓ | ✓ | ✓ | - | - | - | Supervised | - |
| Zhang et al. [12] | - | ✓ | ✓ | ✓ | - | - | - | Supervised | - |
| Ranem et al. [9] | - | ✓ | - | ✓ | - | - | - | Supervised | - |
| Vergani et al. [11] | ✓ | ✓ | ✓ | ✓ | - | - | - | Supervised | - |
| Oksuz et al. [8] | - | ✓ | ✓ | ✓ | - | - | - | Supervised | - |
| Nabavi et al. [16] | ✓ | ✓ | ✓ | ✓ | ✓ | ✓ | ✓ | Unsupervised | 48.73 ± 4.83 |
| Nabavi et al. [18] | ✓ | ✓ | ✓ | ✓ | ✓ | ✓ | ✓ | Meta-Learning | 78.46 ± 6.18 |
| The proposed Model | ✓ | ✓ | ✓ | ✓ | ✓ | ✓ | ✓ | Unsupervised | **84.92 ± 2.53** |

LAX: Long-axis; SAX: Short-axis; CM: Cardiac Motion; RM: Respiratory Motion; GB: Gibbs; AL: Aliasing;

### 4- Discussion

This study attempted to develop an automatic CMR image quality assessment model based on identifying four common types of artefacts. Motion artefacts originating from cardiac and respiratory movements, Gibbs ringing, and Aliasing artefacts were covered in this study. Automatically determining the artefact type can inform the imaging technician about the source of the artefact in real time so that the imaging can be properly repeated. Besides, since the data labelling process is tedious and costly and most of the datasets in practical applications are not annotated, so we focused on the development of an unsupervised model that can address this issue well. This model, with the ability to analyse images and k-space with various views from different distributions, has a suitable generality for evaluating the quality of images in large cohort studies such as UKBB and real-time clinical applications.

The analysis of the results obtained on four different datasets with extensive experiments shows that the proposed model simultaneously has feature discriminability and domain adaptability. By receiving a labelled source set and an unlabeled target set, the proposed model can achieve results very close to the supervised training mode on the labelled target set. This issue confirms that the proposed model can properly address the domain shift problem and lack of access to annotated datasets. The visualisation of the class discrimination and domain adaptation in the source and target sets also reveals this matter. The explainability examination of the model also conveys that this model is focused on the proper area of the image in different classes for decision-making. In detecting cardiac motion artefacts, the model focuses on the heart. In contrast, in detecting lung movements, which resulted in ghosting in the image, the model uniformly concentrates on the ghosting in the entire image. The model detects the Gibbs artefact by paying attention to the areas with visible ringing effects, and this issue is addressed for the aliasing artefact by considering wrap-around areas.

The proposed model, with the possibility of receiving k-space as input, can detect artefacts both in the spatial and frequency domains. Also, the model has been trained on short and long-axis images, which has made the model outperform the other previous methods in terms of covered modality diversity and artefacts. The extensive ablation studies also confirmed the effectiveness of each part of the proposed model.

## 5- Conclusion

In conclusion, deep learning methods suffer performance reduction when trained on a dataset and face data from a different distribution during practical use. A statistical distance-guided unsupervised domain adaptation model was proposed to overcome the domain shift problem, and its performance was evaluated with extensive experiments on several datasets. This model is not a deep network architecture but a general approach to overcome the domain shift in the condition of not having access to the annotated dataset. The model was used to identify four common types of artefacts in CMR imaging. The proposed model can reveal the kind of artefact before further diagnostic analysis by receiving the image or k-space with short or long-axis views post-imaging. Real-time data analysis can lead to the improvement of the diagnosis process. Besides, the model can automatically evaluate the quality of images in large cohort studies. Developing the model to enable multiple target sets domain adaptation can be a potential future study.

## Acknowledgements

This research has been done using the UK Biobank database under Application 11350.

## Statements and Declarations

There are no conflicts of interest to declare. No funding was received for conducting this study.

**Supplementary Materials**

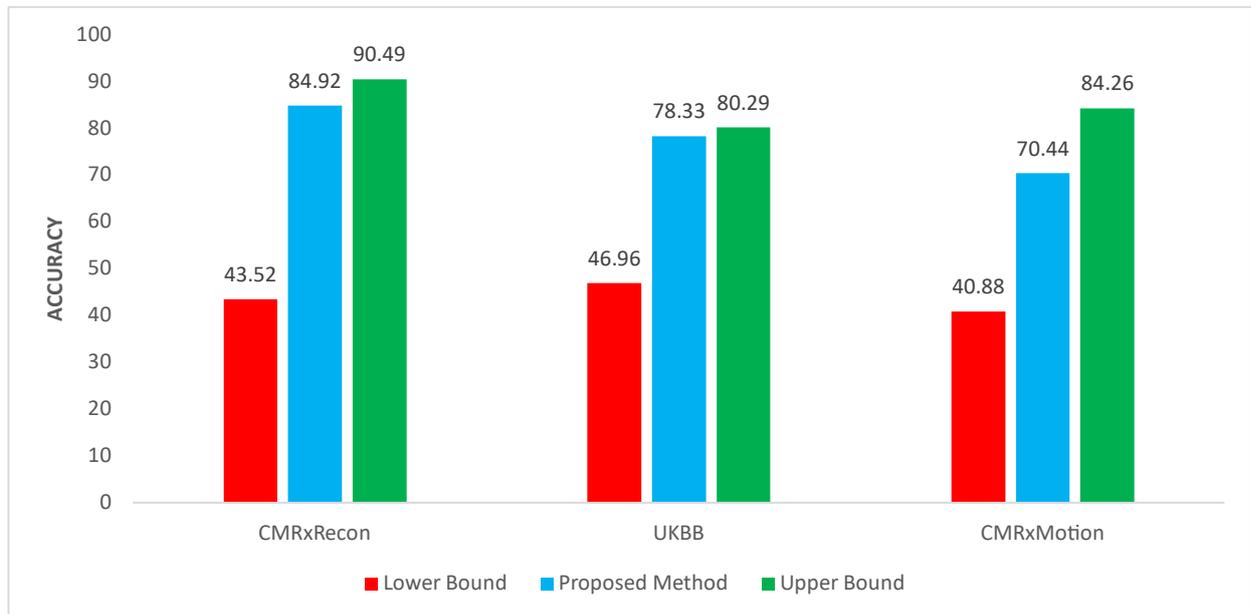

**Figure S1:** The domain gap coverage using the proposed model, considering ACDC as a source set and CMRxRecon, UKBB, and CMRxMotion as target sets. (Lower bound: Train on source/Test on target; Upper bound: Train on target/Test on target)

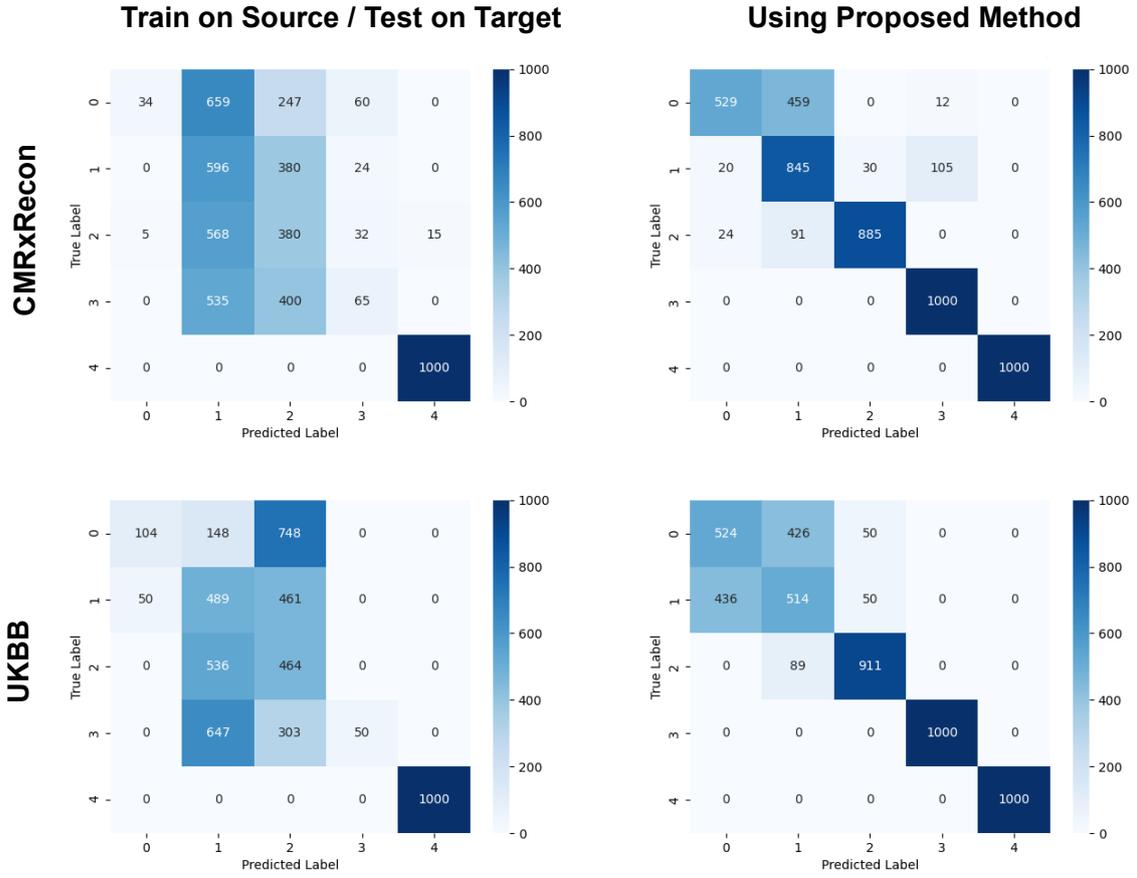

**Figure S2:** Confusion matrices related to the 5-class artefact classification before and after using the proposed model, considering ACDC as the source set and UKBB and CMRxRecon as the target sets.

**Tabel S1: The results of using the proposed model on the k-space data.**

| Dataset | Metric | Accuracy | Precision | Recall | F1-Score | AUC |
|---|---|---|---|---|---|---|
| Source: CMRxRecon  Target: UKBB | Train on source/Test on target | 44.24 ± 5.73 | 37.93 ± 8.57 | 44.24 ± 5.73 | 36.38 ± 5.57 | 76.99 ± 4.61 |
| | Using the proposed model | 66.24 ± 5.09 | 74.24 ± 2.60 | 66.24 ± 5.09 | 61.58 ± 6.37 | 88.04 ± 3.35 |
| | Train on target/Test on target | 70.91 ± 6.41 | 74.38 ± 1.11 | 70.91 ± 6.41 | 68.25 ± 8.49 | 90.32 ± 1.62 |